# Collective Strategies with a Master-slave Mechanism Dominate in Spatial Iterated Prisoner's Dilemma

Jiawei Li[1], Robert Duncan[2], Jingpeng Li[2], Ruibin Bai[1]

1 School of Computer Science, University of Nottingham Ningbo, China

2 Department of Computing Science & Mathematics, University of Stirling, UK

**Abstract:** How cooperation emerges and persists in a population of selfish agents is a fundamental question in evolutionary game theory. Our research shows that Collective Strategies with Master-Slave Mechanism (CSMSM) defeat Tit-for-Tat and other well-known strategies in spatial iterated prisoner's dilemma. A CSMSM identifies kin members by means of a handshaking mechanism. If the opponent is identified as non-kin, a CSMSM will always defect. Once two CSMSMs meet, they play master and slave roles. A mater defects and a slave cooperates in order to maximize the master's payoff. CSMSM outperforms non-collective strategies in spatial IPD even if there is only a small cluster of CSMSMs in the population. The existence and performance of CSMSM in spatial iterated prisoner's dilemma suggests that cooperation first appears and persists in a group of collective agents.

## 1. Introduction

The Prisoner's Dilemma is a two-player non-zero-sum game in which two players try to maximize their payoffs by cooperating with or betraying the other player. In the classical version of prisoner's dilemma, each player chooses between two strategies, Cooperate (C) and Defect (D). Their payoffs can be represented by the matrix shown in Figure 1.

| | | ***Player II*** | |
|---|---|---|---|
| | | Cooperate | Defect |
| ***Player I*** | Cooperate | ***R, R*** | ***S, T*** |
| | Defect | ***T, S*** | ***P, P*** |

Fig. 1. Payoff matrix of the Prisoner's Dilemma.

In the payoff matrix, $R$, $S$, $T$, and $P$ denote Reward for mutual cooperation, Sucker's payoff, Temptation to defect, and Punishment for mutual defection respectively, and $T > R > P > S$. The constraint motivates each player to play non-cooperatively.

When both players are rational and they make their choice independently, the theoretical outcome of the game is a Nash equilibrium, in which both players choose to defect, and each receives a 'Punishment for mutual defection'. It is worse for each player than the outcome they would have received if they had cooperated. [1,2]

In the Iterated Prisoner's Dilemma (IPD), two players have to choose their mutual strategy repeatedly, and they also have memory of their previous behaviors and the behaviors of the opponents. There is $R > 1\ 2(S + T)$, which is set to prevent any incentive to alternate between cooperation and defection. IPD is considered to be an ideal experimental platform for the evolution of cooperation among selfish individuals and it has attracted wide interest since Robert Axelrod's IPD tournaments and his book 'The Evolution of Cooperation'[3].

If the precise length of an IPD is known to the players, the best strategy for both players is to defect in each move. This is a conclusion from backward induction: both players will choose to defect in the final iteration because the opponent will not be able to subsequently punish the player. Given mutual defection in the final iteration, the optimal strategy in the penultimate iteration is defection for both players, and so on, back to the initial iteration. If the precise length of an IPD is infinite or unknown, mutual cooperation can also be equilibrium.

Axelrod was the first to study efficient IPD strategies by means of competitions [4, 5]. TFT always cooperates in the first move and then mimics whatever the opponent did in the previous move. According to Axelrod, several characteristics make TFT successful: TFT is Nice, Retaliating and Forgiving. TFT is not a Nash equilibrium and there is always a sub-game perfect equilibrium that dominates TFT, according to the Folk Theorem in game theory [6, 7]. On the other hand, whether or not TFT is the most efficient strategy in IPD is still unclear. Some strategies perform better than TFT in specific environments [8-12]. Therefore, researchers are attempting to develop novel strategies that can outperform TFT either in round-robin tournaments or in evolutionary dynamics.

In recent IPD competitions, strategies have appeared with identification mechanisms. With a rule-based identification mechanism, a strategy called APavlov won competition four of the 2005 IPD competition [13]. Furthermore, many of the top listed strategies somehow explore the opponent by using simple mechanisms [14, 15]. This shows that strategies that *explore and then exploit* the opponent can outperform any single non-group strategy in round-robin IPD tournaments.

A strategy with a simple identification mechanism, named 'handshake' [16], appeared in evolutionary IPD. This strategy defects at the first move and cooperates on the second move. If the opponent behaves in the same way, it will keep cooperating in the following moves. Otherwise, it will always defect. This 'initial defect then cooperate' can be seen as a password. Any strategy that knows this password (or behaves the same by chance) may evoke handshake's cooperation while other actions trigger defection. By means of a mechanism like 'handshake', a group of strategies are able to recognize each other and then behave collectively [17, 18].

In the 2004 IPD competition, a team from Southampton University led by Jennings introduced a group of strategies, which outperformed all singleton strategies, and won the top three positions. These strategies were designed to recognize each other through a predetermined sequence of 5-10 moves at the start. Once two Southampton players recognized each other, they would act as a 'master' or 'slave' - a master will always defect while a slave will always cooperate in order for the master to win the maximum points. If the opponent was recognized as not being a Southampton entry, it would immediately defect to minimize the score of the opponent [19]. The Southampton strategies were designed to maximize the payoffs of a small number of masters so that they could win in a round robin IPD tournament. The slaves performed poorly and they generally received less payoffs than their opponents. This master-slave scheme is ineffective in evolutionary dynamics because the slaves quickly die out and then the masters lose the advantage of exploiting the slaves.

In this paper, we show that a novel strategy, so-called Collective Strategy with Master-Slave Mechanism (CSMSM), dominates in spatial IPD. CSMSM outperforms other IPD strategies when they form clusters. Simulation results show that CSMSM always outperforms non-collective opponents once they form a cluster in the initial population.

Existence of strategies like CSMSM suggests that cooperation first appears in a group of agents who cooperates with group members and defects against non-members. Once those agents form a small cluster, they outperform defective neighbours and their percentage grows steadily in the population.

## 2. Collective strategies with a master-slave mechanism

Based on handshaking, a group of strategies can recognize each other and deal with group members and non-members differently. A handshake in IPD is a predefined sequence of moves for two interacting players. If both players play the same sequence of moves, they recognize each other as group members. Otherwise, they identify the opponent as non-kin.

In this paper, the handshake of CSMSM is a sequence of 'CDCCD'. When two CSMSMs meet, they will play this sequence in the first five moves and then recognize each other as a group member. If the opponent has played a different sequence, this immediately triggers defection and a CSMSM will always defect against the opponent from that point on.

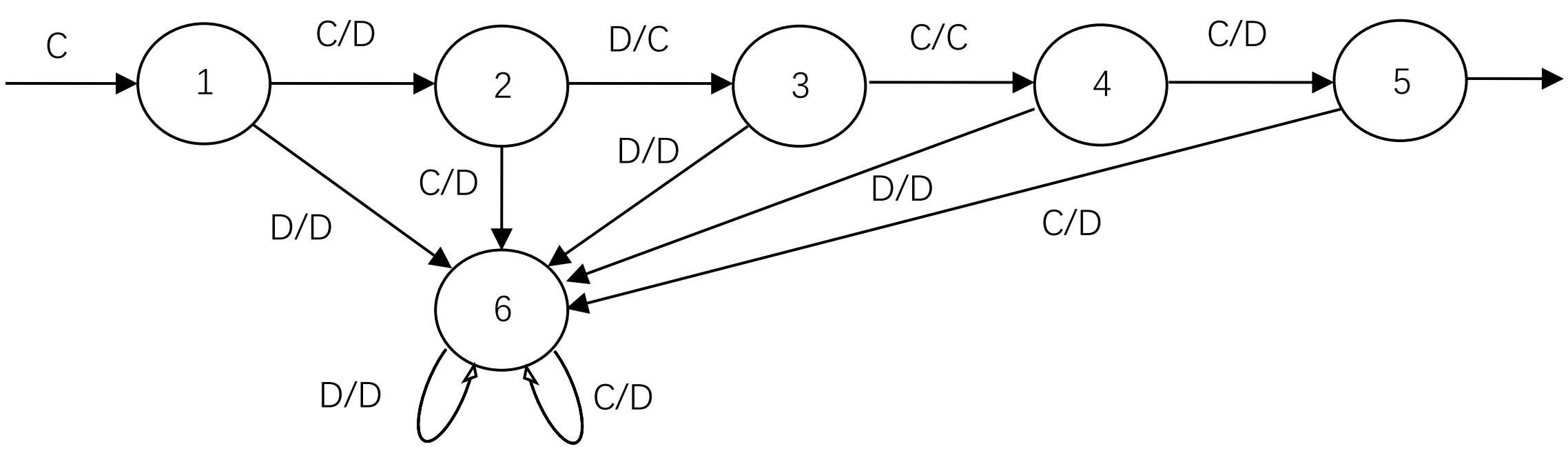

Fig. 2 Moore finite machine expression of the handshake mechanism of the CSMSM strategy Payoff.

As long as two CSMSMs have identified each other, they will play either a master or a slave role. Every CSMSM behaves as a master originally. In each generation, a master CSMSM may change to a slave with a predefined probability (it is set to be 0.7 in this study), which makes sure that there is always a ratio of slaves in the CSMSM group. Because the slaves are designed to sacrifice them in order to maximize the masters' payoffs, they are likely to receive lower payoff than the average of the population and then they die out in each generation. Thus, the percentage of slaves in CSMSM will be approximately 70% when the probability of a master changing to a slave is set to be 0.7. A slave will always be a slave. The impact of the ratio of slaves on the performance of CSMSM will be discussed later.

As long as two CSMSMs have identified each other, a master will act as a Grim trigger (it chooses 'C' as long as the opponent chooses 'C'. Once the opponent has chosen 'D', it always defects) and a slave will act as a reverse TFT (it chooses 'D' first and then chooses the reverse of the opponent's previous choice). When a master and a slave meet, the master chooses 'C' and the slave chooses ‘D’ in the sixth move, and then the master will always play ‘D’ and the slave will always play ‘C’ so that the master’s payoff in the following rounds is maximized.

When two masters meet, they will cooperate with each other except handshaking in the first five moves. When two slaves meet, they will play a sequence of alternate defection and cooperation after handshaking.

As an example, let’s consider how a CSMSM interacts with a TFT. Both strategies play ‘C’ in the first move. Then TFT plays ‘C’ and CSMSM plays ‘D’ in the second move. Since TFT has not played the handshaking sequence, CSMSM will play ‘D’ thereafter. They will keep defecting after the third move.

When CSMSM forms a cluster in the population, they outperform any individual strategy. In the next section, we analyse the performance of CSMSM in spatial IPD game.

## 3. CSMSM in spatial IPD

As a collective strategy, the performance of CSMSM in a spatial IPD depends on the size of the clusters they form. Whenever CSMSM has formed a cluster of significant size, they will outperform their competitors.

### 3.1 CSMSM vs. TFT

Let’s first consider a typical scenario that a cluster of nine CSMSM is surrounded by TFTs, as shown in Fig.3a. The minimum size of a cluster for CSMSM to survive in a population of TFT is nine. Let $n$ ($n$>6) denote the number of iterations in the IPD.

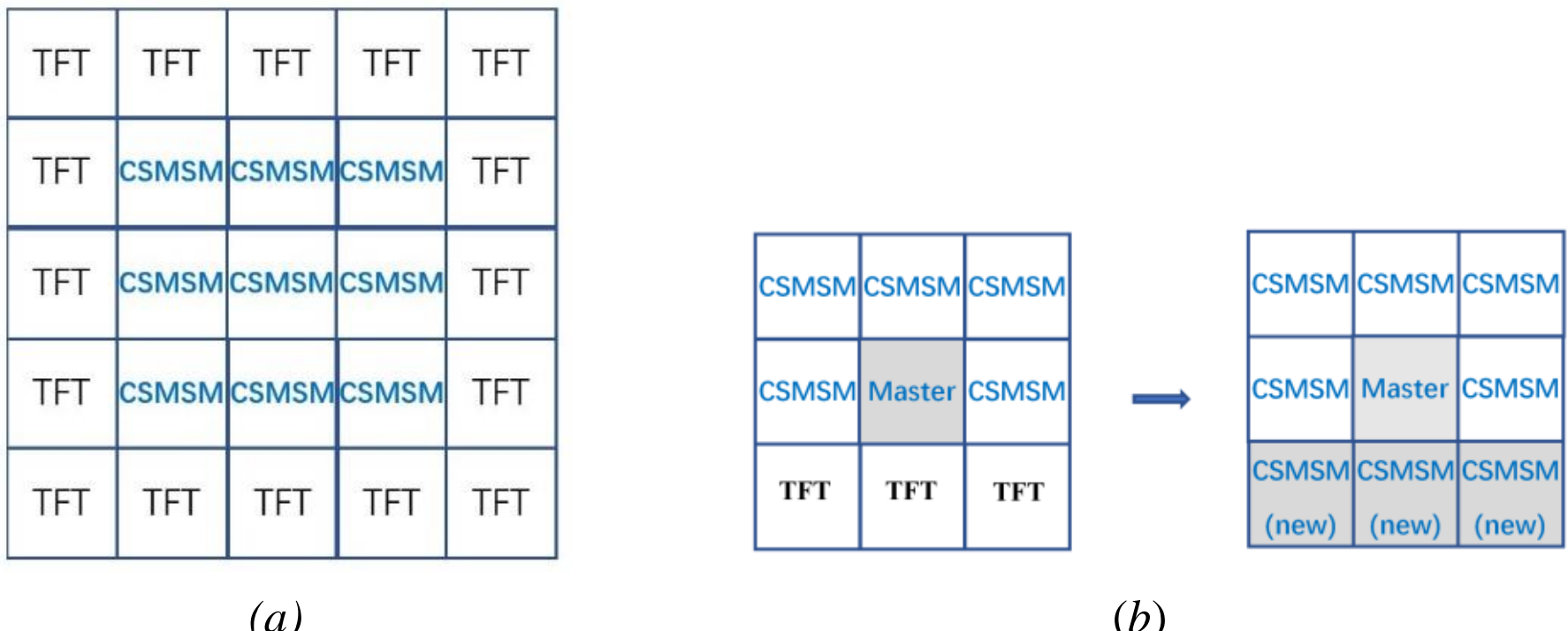

(*a*) (*b*)

Figure 3 (*a*) a cluster of nine CSMSMs is surrounded by a population of TFT. (*b*) A master on border who has four or five slaves as direct neighbours receives the highest payoff in the local area, and thus it will invade the TFT territory.

A TFT receives $R+(n-2)P+S$ in interacting with a CSMSM and receives $nR$ against another TFT. A master CSMSM receives $T+R+(n-2)P$ against a TFT, $(n-2)R+2P$ against a master, and $(n-6)T+3R+2P+S$ against a slave.

Without self-interaction, each strategy plays eight IPDs with its direct neighbours. A TFT may have zero, one, two, or three CSMSM neighbours depending on its position. With zero CSMSM and eight TFT neighbours, a TFT receives $T_0 = 8nR$ payoff. With one, two, or three CSMSM neighbours, the payoff of a TFT is $T_1 = 7nR + (R+(n-2)P+S)$, $T_2 = 6nR + 2\times(R+(n-2)P+S)$, or $T_3 = 5nR + 3\times(R+(n-2)P+S)$ respectively. There are $T_1 > T_2 > T_3$ since

$R > P > S$.

For CSMSM, we only need to consider the payoffs of the masters because the slaves generally receive lower payoffs than TFTs. Depending on the position of a master, it may have three, five, or eight CSMSM neighbours.

(***a***) Firstly, let's consider the scenario that the centre of the cluster of nine CSMSM is a master. Let *m* ($8 \geq m \geq 0$) denote the number of slaves in the cluster. The payoff of the master can be computed by $C_0 = m((n-6)T+3R+2P+S)+(8-m)((n-2)R+2P)$. We have

$$C_0 > T_1 \tag{1}$$

when $n > \frac{17R-18P+S}{R-P}$ and whatever value of *m* is. The payoff of the central master is the highest within the 5x5 area when the number of iterations is long enough, for example $n > 16$ when $R$=3, $P$=1, $S$=0. Thus, the cluster of CSMSM cannot be invaded by TFT if the central strategy is a master.

(***b***) Secondly, let's consider the scenario that a master is located on one side border of the cluster (with five CSMSM neighbours). Let *l* ($5 \geq l \geq 0$) denote the number of slaves in the five CSMSM neighbours. The payoff can be computed as $C_1 = l((n-6)T+3R+2P+S)+(5-l)((n-2)R+2P)+3\times(T+R+(n-2)P)$. In order that $C_1 > T_0$, we have

$$l > \frac{3n(R-P)-3T+7R-4P}{n(T-R)-6T+5R+S} \tag{2}$$

In the case that $T$=5, $R$=3, $P$=1, $S$=0 and $n$=50, (2) can be computed as $l > 3.55$. It means that the master receives greater payoff than the highest payoff of TFT if $l$=4 or 5. The three TFT neighbours will be replaced by CSMSM, as shown in Fig. 3b.

In the case that *n* is far greater than *T*, $n$>50 for example, (2) can be simplified to (3).

$$l > \frac{3(R-P)}{T-R} \tag{3}$$

Under the condition of (2) or (3), the cluster of CSMSM grows in size.

In the case that $T_0 > C_1 > T_1$, we have,

$$\frac{3n(R-P)-3T+7R-4P}{n(T-R)-6T+5R+S} > l > \frac{2n(R-P)-3T+8R-6P+S}{n(T-R)-6T+5R+S} \tag{4}$$

It can be simplified to (5) when *n* is a relatively large value.

$$\frac{3(R-P)}{T-R} > l > \frac{2(R-P)}{T-R} \tag{5}$$

Under the condition of (4) or (5), the payoff of the master is higher than its TFT neighbours and lower than the highest payoff of TFT. Thus, the cluster of CSMSM will maintain the current size.

(***c***) Thirdly, in the scenario that a master locates on one corner of the cluster (with three

CSMSM neighbours), the payoff is $C_2 = q((n\text{-}6)T+3R+2P+S)+(3\text{-}q)((n\text{-}2)R+2P)+5\times(T+R+(n\text{-}2)P)$, where $q$ ($3 \geq q \geq 0$) is the number of slaves in the three CSMSM neighbours. It can be verified that $C_2$ is always smaller than $T_2$ . Thus, a master on the corner always receives lower payoff than its TFT neighbours.

In summary, if there is a master on the border and the number of its slave neighbours satisfies (2) or (3), the neighbour TFTs will be replaced by CSMSM and the cluster of CSMSM grows. Otherwise, if the centre of the cluster is a master, or there is a master on the border and (4) or (5) satisfies, the cluster will maintain the current size. In other situations, the cluster of CSMSM may be invaded by TFT.

In the case that $T$=5, $R$=3, $P$=1, $S$=0 and $n$=50, the ratio of slaves in CSMSM should be at least 30% so that there are approximately three slaves in the neighbourhood of a master on border, which makes sure that the border cannot be invaded by TFT. If the ratio of slaves can be maintained higher than 30%, CSMSM outperforms TFT in spatial IPD.

The result of a simulation of CSMSM competing against TFT is shown in Fig.4, where the initial population is 50% TFT and 50% CSMSM, randomly allocated on a 200×200 grid. The changes of the percentages of two strategies in the population and average payoffs are illustrated in Fig. 5(a) and (b) respectively. It shows that the masters outperform TFT decreases rapidly and it dies out after 13 generations.

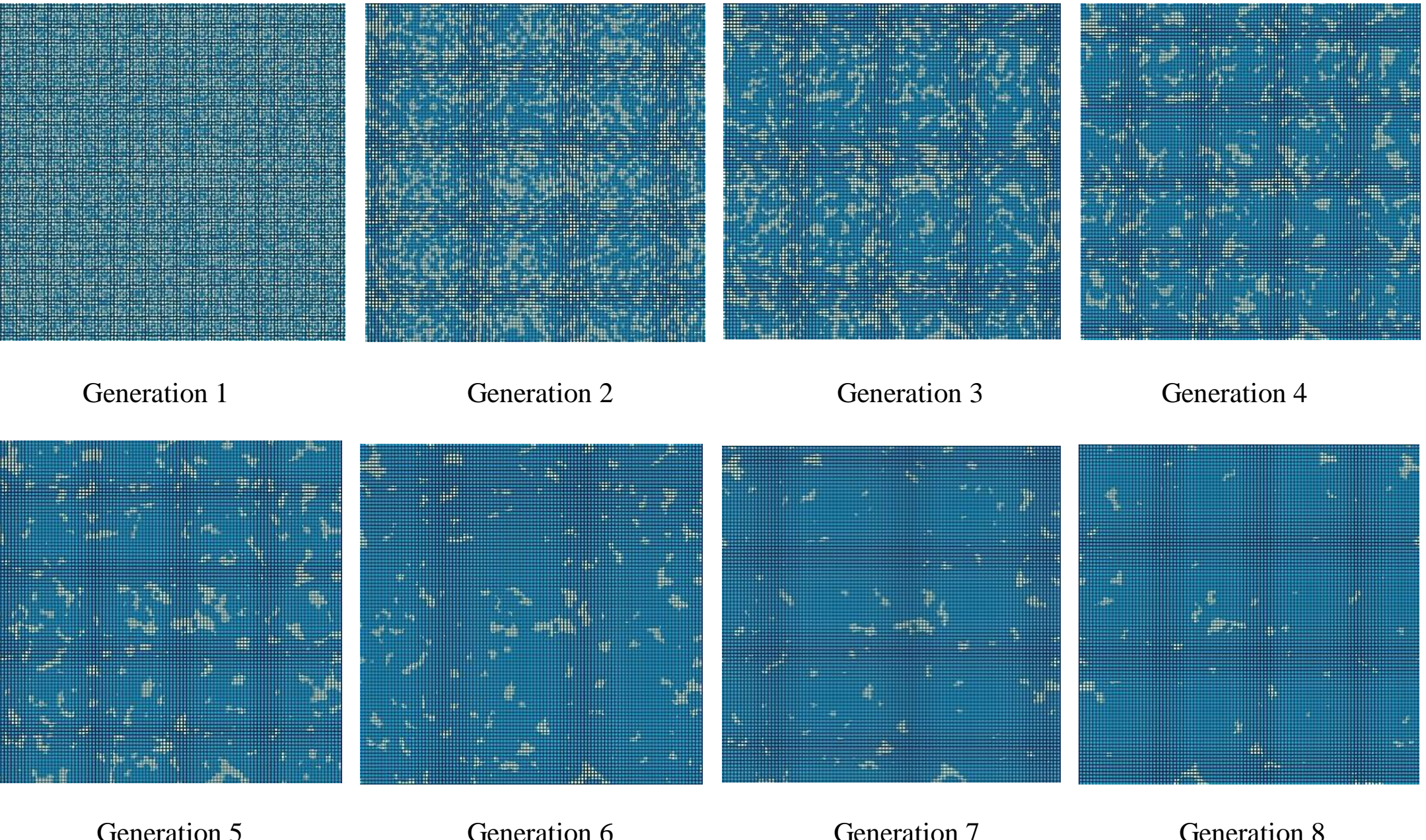

Figure 4 CSMSM defeats TFT in a spatial IPD. The initial population contains 50% CSMSM and 50% TFT. The dark colour squares denote CSMSM while the light colour denotes TFT.

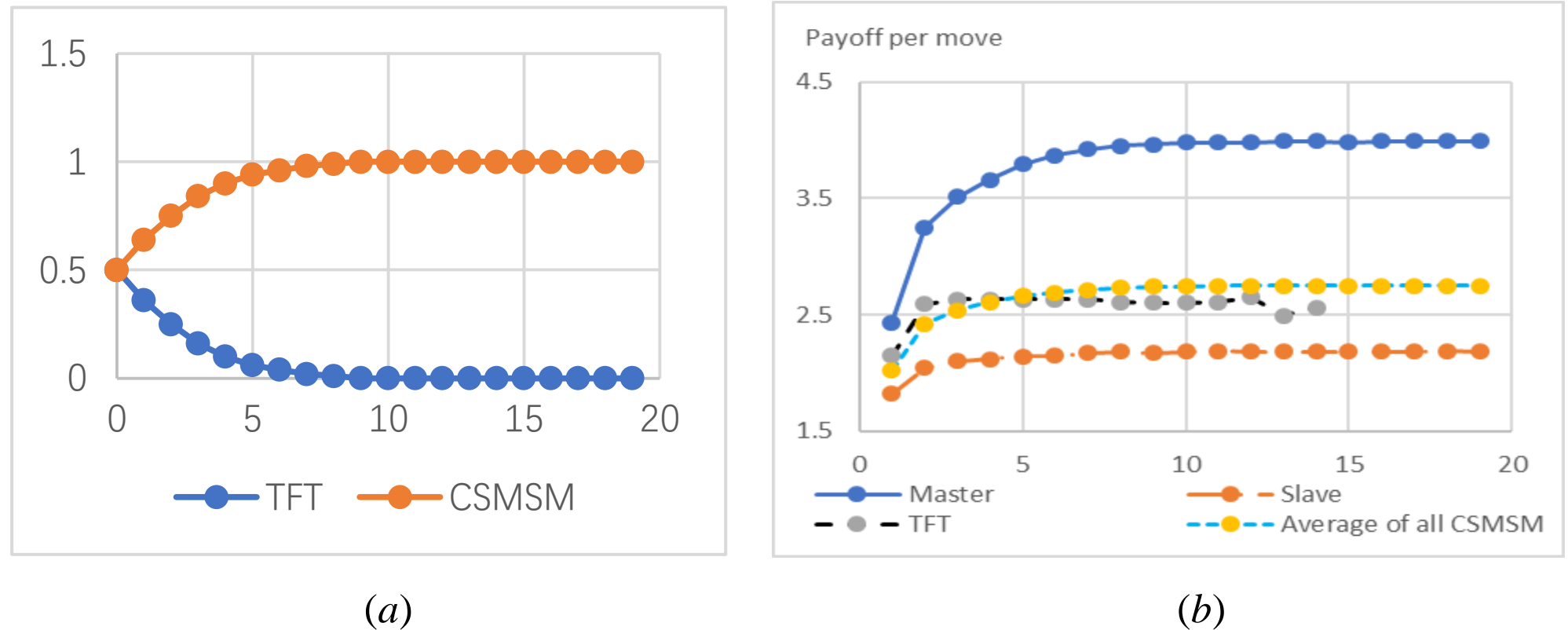

(*a*) (*b*)

Figure 5 (*a*) The percentage of CSMSM and TFT. The x-axis denotes generation and y-axis denotes the ratio of strategies in the population. (*b*) The average payoff per move of masters, slaves, TFT, and all CSMSM.

### 3.2 CSMSM vs. AllD

We now consider another scenario that a cluster of two CSMSM is surrounded by AllDs, as shown in Fig. 6a. Each CSMSM has seven AllD and one CSMSM neighbours. An AllD may have zero, one, or two CSMSM neighbours.

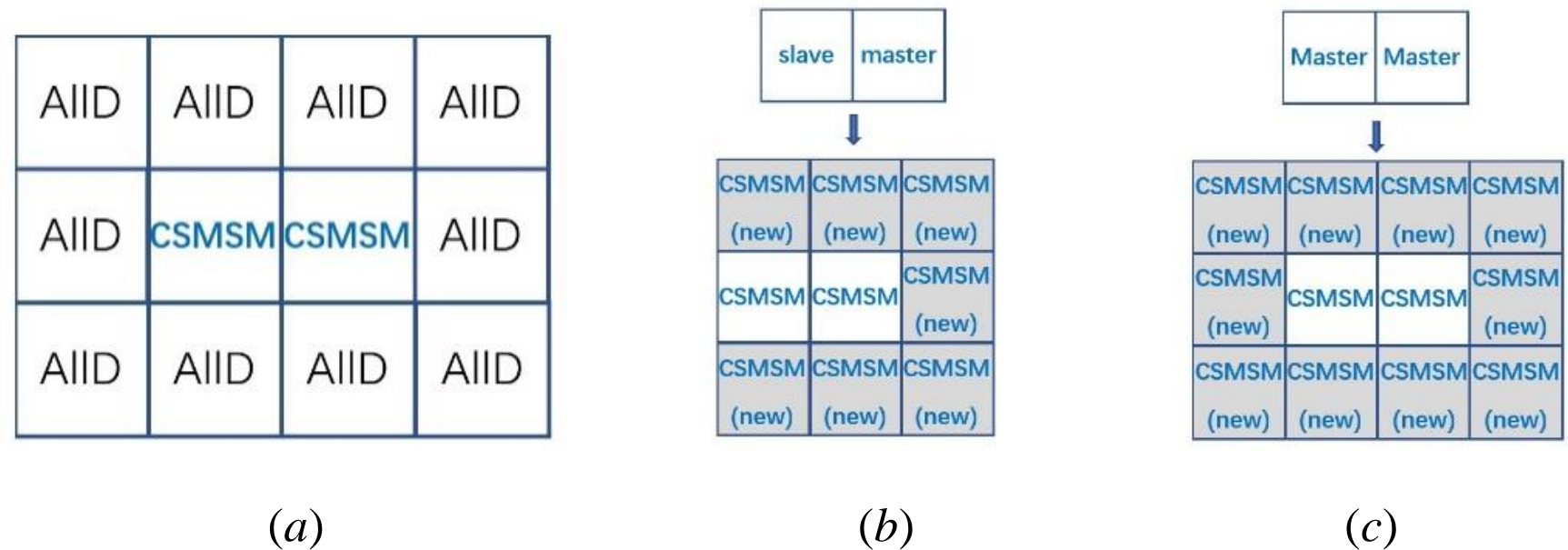

(*a*) (*b*) (*c*)

Figure 6 (*a*) a cluster of two CSMSMs is surrounded by a population of AllD. (*b*) If they are one master and one slave, the master receives higher payoff than surrounding AllDs and thus it will invade the AllD territory. (*c*) If they are two masters (or two slaves), the masters (or slaves) receive higher payoff than surrounding AllD.

An AllD receives $T+(n-1)P$ in interacting with a CSMSM and receives $nP$ against another AllD. A master CSMSM receives $(n-1)P+S$ against an AllD, $(n-2)R+2P$ against a master, and $(n-6)T+3R+2P+S$ against a slave. A slave receives $(n-1)P+S$ against an AllD, $T+3R+2P+(n-6)S$ against a master, and $(nR+nP)/2$ against a slave.

With zero, one, and two CSMSM neighbours, the payoff of an AllD are $8nP$, $7nP+(T+(n-1)P)$, and $6nP+2\times(T+(n-1)P)$ respectively. The greatest of three values is $D_2 = 8nP+2T-2P$.

If two CSMSMs are a master and a slave, the master receives $7\times((n-1)P+S)+((n-6)T+3R+2P+S)$ and the slave receives $7\times((n-1)P+S)+(T+3R+2P+(n-6)S)$. The payoff of the master is higher than $D_2$, the highest payoff of AllD, when there is

$$n > \frac{8T-3R+3P-2S}{T-P} \tag{6}$$

In the case that $T$=5, $R$=3, $P$=1, $S$=0 for example, there is $n>8.5$. Under this condition, the AllD neighbours around the master will be replaced by CSMSM, as shown in Fig. 4b.

If both CSMSMs are masters, their payoff is $7\times((n-1)P+S)+((n-2)R+2P)$ and it is higher than $D_2$ when (6) is satisfied. If both CSMSMs are slaves, their payoff is $7\times((n-1)P+S)+(nR+nP)/2$ and it is higher than $D_2$ as well. Thus, the AllDs in the neighbourhood of the two CSMSMs will be replaced, as shown in Fig. 4c. Thus, a cluster of two or more CSMSM outperforms AllD in spatial IPD.

### 3.3 CSMSM invades a population of another strategy

We have simulated CSMSM competing against some well-known strategies in spatial IPD. The results are given in Fig. 7. The initial population in each simulation contains 20% CSMSM and 80% another strategy, randomly mixed and allocated on a 200×200 grid. In all simulations, CSMSM outperforms the opponent strategy in 200 generations.

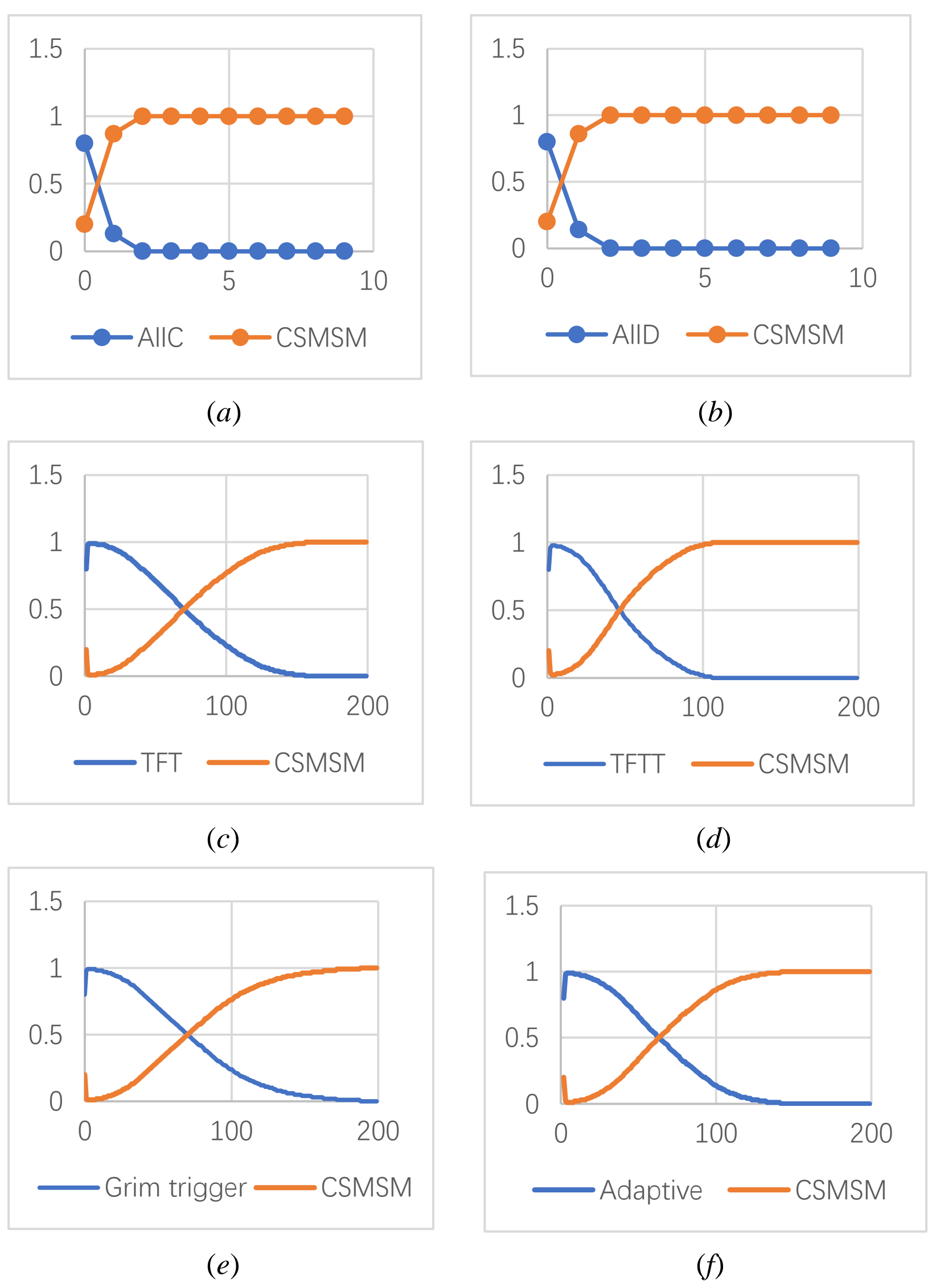

Figure 7 CSMSM outperforms some well-known strategies. The initial population is 20% CSMSM and 80% another strategy. The x-axis denotes generation and y-axis denotes the ratio of strategies in the population. (*a*) CSMSM vs. AllC. (*b*) CSMSM vs. AllD. (*c*) CSMSM vs. TFT. (*d*) CSMSM vs. TFTT.

TFTT: Tit for two tat. (*e*) CSMSM vs. Grim trigger. (*f*) CSMSM vs. Adaptive. Adaptive: it starts with C,C,C,C,C,C,D,D,D,D,D and then takes choices which have given the best average score re-calculated after every move.

As an example, the process of competition between CSMSM and TFT in a spatial IPD is shown in Fig. 8. The numbers of CSMSM decreases sharply at the beginning stage because they have not formed clusters. Once small clusters appear, their percentage starts to increase. When CSMSM has formed clusters of significant sizes, they grow steadily in size and expel TFT as a result.

For some other strategies like TFTT, Grim trigger, and Adaptive, similar process as Fig. 8 can be observed.

For some naive strategies like AllC, AllD, and Random player, CSMSM expels them in 3-5 generations because it does not need clusters of significant size to invade the territory of those strategies.

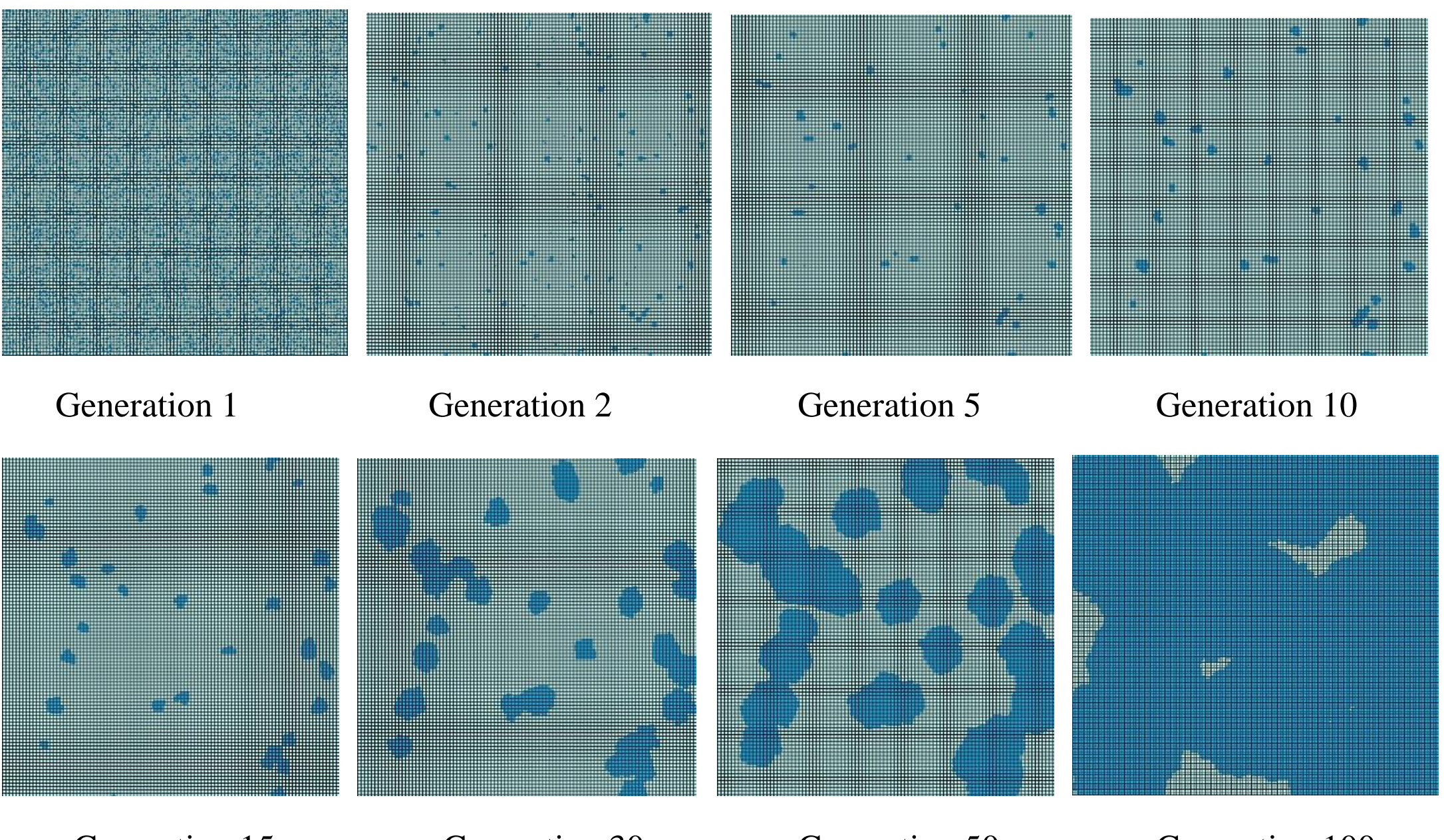

Figure 8 CSMSM outperforms TFT when the initial population is 20% CSMSM and 80% TFT. The dark colour squares denote CSMSM while the light colour denotes TFT.

## 4. Discussion and Conclusions

We have presented a CSMSM strategy that outperforms well-known strategies such as TFT in spatial IPD. By means of a handshaking and a master-slave mechanism, CSMSM manages to maximize the masters' payoffs and minimize the opponents' payoffs. Spatial IPD simulations show that CSMSM outperforms other strategies even if it is minority in the initial population. Once CSMSM has formed a cluster of significant size, it will dominate in the evolution as a result.

The replicator equation (7) is the most commonly used model in evolutionary game theory.

$$\dot{x}_i = x_i[f_i(x) - \phi(x)],\ \phi(x) = \sum_{j=1}^{n} x_j f_j(x) \quad (7)$$

It has been revealed that a stable rest point of the equation is also a Nash equilibrium [32], so stable states in evolutionary games can be predicted by means of the folk theorems [33]. Our research shows that if an evolutionary process is not as simple as the replicator equation models, the stable states are not necessarily Nash equilibria.

The master-slave mechanism and handshaking have been neglected in game theory research. One reason may be that they do not follow the assumption of self-interest players, so they cannot be well explained by means of Nash equilibrium. Handshaking provides a method for those players who have the same objective to recognize each other and to coordinate their choices. The master-slave mechanism provides a method for the same strategies to win in a winner-take-all environment. There is still a need for a theory for these kinds of behaviours.

The master-slave mechanism specially fits for winner-take-all situations. It is economically reasonable only when the winning of a master is enough to compensate the loss of the slaves who are sacrificed. In a winner-take-all situation, a master-slave strategy profile can be profitable for a group of individuals, especially in an evolutionary context.

Note that CSMSM never coexists with any other strategy unless that strategy behaves in the same way as a CSMSM. This feature makes CSMSM especially strong in maintaining a homogeneous population in evolution. A game theoretical model that provides explanations for interactive strategies like CSMSM will be our future research.